\documentclass{aa}
\usepackage{txfonts}
\usepackage{graphicx}
\usepackage[section]{placeins}
\usepackage{multirow}
\usepackage{amsmath}
\usepackage{xcolor}

\newcommand{\dScu}{{$\delta$~Scuti }}
\newcommand{\gDor}{{$\gamma$~Doradus }}

\bibpunct{(}{)}{;}{a}{}{,}

\begin{document}

\title{Unveiling the power spectra of \dScu stars with \textit{TESS}}
\subtitle{The temperature, gravity, and frequency scaling relation}

\author{S.~{Barcel\'o Forteza}\inst{\ref{ins0}} \and A.~{Moya}\inst{\ref{ins1},\ref{ins2}} \and D.~{Barrado}\inst{\ref{ins0}} \and E.~{Solano}\inst{\ref{ins0},\ref{ins25}} \and S.~{Mart\'in-Ruiz}\inst{\ref{ins3}}  \and J.~C.~{Su\'arez}\inst{\ref{ins4},\ref{ins3}} \and A.~{Garc\'ia~Hern\'andez}\inst{\ref{ins4},\ref{ins3}}}

\institute{Dpto. de Astrof\'isica, Centro de Astrobiolog\'ia (CSIC-INTA), ESAC, Camino Bajo del Castillo s/n, 28692, Spain \label{ins0}
\and 
Electrical Engineering, Electronics, Automation and Applied Physics Department, E.T.S.I.D.I, Polytechnic University of Madrid (UPM), Madrid 28012, Spain \label{ins1}
\and 
School of Physics and Astronomy, University of Birmingham, B15 2TT, UK. \label{ins2}
\and
Spanish Virtual Observatory, Spain\label{ins25}
\and
Instituto de Astrof\'isica de Andaluc\'ia (CSIC), Glorieta de la Astronom\'ia s/n, 18008, Granada, Spain\label{ins3}
\and
Dept. Theoretical Physics and Cosmology, University of Granada (UGR), 18071, Granada, Spain \label{ins4}
}

\date{Received 5 December 2019; Accepted 14 April 2020}

\abstract{Thanks to high-precision photometric data legacy from space telescopes like \textit{CoRoT} and \textit{Kepler}, the scientific community could detect and characterize the power spectra of hundreds of thousands of stars. Using the scaling relations, it is possible to estimate masses and radii for solar-type pulsators. However, these stars are not the only kind of stellar objects that follow these rules: \dScu stars seem to be characterized with seismic indexes such as the large separation ($\Delta\nu$). Thanks to long-duration high-cadence \textit{TESS} light curves, we analysed more than two thousand of this kind of classical pulsators. In that way, we propose the frequency at maximum power ($\nu_{\rm max}$) as a proper seismic index since it is directly related with the intrinsic temperature, mass and radius of the star. This parameter seems not to be affected by rotation, inclination, extinction or resonances, with the exception of the evolution of the stellar parameters. Furthermore, we can constrain rotation and inclination using the departure of temperature produced by the gravity-darkening effect. This is especially feasible for fast rotators as most of \dScu stars seem to be.}


\maketitle

\section{Introduction}
\label{s:intro}

Asteroseismology has proven to be a very fruitful technique to characterize the stars and improve the stellar evolution theory \citep[see][for detailed reviews]{Aerts2010,Chaplin2013,Catelan2015,Aerts2019}. Stellar pulsations are sensitive to the internal structure of stars and their physics. There are different types of pulsators (e.g. solar-like, \dScu stars) according to their excitation mechanism \citep[see Fig.~1 in][]{Jeffery2008}. The seismic indexes describe the properties of the power-spectral structure as a whole, the so-called power spectrum envelope (envelope hereafter). The scaling relations can relate their structural parameters with seismic indexes such as happens with solar-like pulsators \citep[e.g.,][]{Kjeldsen1995}. Using the high amount of stellar data from space missions like \textit{CoRoT} \citep{Baglin2006} or \textit{Kepler} \citep{Borucki2010}, the scientific community has been looking for scaling relations also for \dScu stars \citep[e.g.,][]{Suarez2014a,GarciaHernandez2015,Michel2017,Moya2017,BarceloForteza2018,Bowman2018}. Trying to extend ensemble asteroseismology, \citet[][-BF18 hereafter-]{BarceloForteza2018} describe the envelope of \dScu stars with several metrics such as the number of modes ($N_{\rm env}$), the frequency at maximum power, 
\begin{equation}
\nu_{\rm max} = \frac{\sum A_{i}\nu_{i}}{\sum A_{i}} \, ,
\label{e:numax}
\end{equation}
where $\nu_{i}$ and $A_{i}$ are the frequency and the amplitude of each mode of the envelope, respectively; and its asymmetry
\begin{equation}
\alpha = \frac{2\nu_{\rm max} - \nu_{h} - \nu_{l}}{2 \left(\nu_{h} - \nu_{l} \right)} \, ,
\label{e:alph}
\end{equation}
where $\nu_{h/l}$ are the highest and lowest frequency of the envelope, respectively.

\begin{figure}
\includegraphics[width=\linewidth]{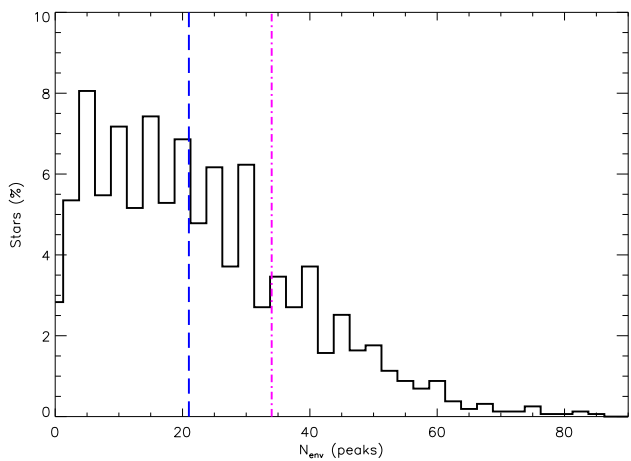}
\caption{Distribution of stars according to the number of peaks in their envelope. Dashed blue line points to the mean number of peaks of the envelope and the purple dashed-dotted to the estimated by \cite{Lignieres2009}.}
\label{f:nenv}
\end{figure}

\dScu stars are A-F intermediate mass stars \citep[1.5 to 2.5 $M_{\sun}$;][]{Breger2000} with frequencies between 60 to 930 $\mu$Hz and temperatures from 6000 to 9000 K \citep{Uytterhoeven2011}. Their main excitation mechanism is $\kappa$-mechanism \citep{Chevalier1971}. \cite{Dziembowski1997} predicted that the excited modes have higher frequencies at higher temperatures ($T_{\rm eff} \propto \nu_{i}$, see Fig.~2 in that paper). Taking into account solar composition, but no rotation, and no core overshoot, \cite{Balona2011} also predicted this behaviour for the frequency of the mode with highest amplitude, 
\begin{equation}
T_{\rm eff} \propto \nu_{0} \,.
\label{e:sr1}
\end{equation}
However, the observations show a wide variation (see Fig.~2 in that paper). These differences may be produced by other mechanisms playing a significant role, especially for hybrid pulsators \citep{Antoci2014,Xiong2016}. On the other hand, there are other physical processes that can modify the observed temperature such as the gravity-darkening effect \citep{vonZeipel1924}. A high rotation rate modifies the shape of the star from a sphere to an ellipsoid. In that way, the temperature at the poles is higher than the temperature at the equator. The departure of temperature is defined by BF18 as
\begin{equation}
\delta \bar{T}_{\rm eff}(i) \equiv \frac{T_{\rm eff}(i)-\bar{T}_{\rm eff}}{\bar{T}_{\rm eff}} \approx \left( \frac{1 - \frac{R(i)}{R}\epsilon^2 \sin^2\{i\}}{1-\frac{2}{3}\epsilon^2}\right)^\frac{\beta}{4} -1
\label{e:gde}
\end{equation}
where $i$ is the inclination from the line of sight; $\beta$ depends on the importance of the convection \citep{Claret1998}; and $\epsilon$ is the ratio between the centrifugal and gravity forces
\begin{equation}
\epsilon^2 = \frac{\Omega^2 R^3}{GM}\, \,
\label{e:eps2}
\end{equation}
where $M$ is the mass; $\bar{T}_{\rm eff}$ and $R$ are the mean effective temperature and the mean radius, i.e., the temperature and radius of a spherically symmetric star with the same mass as the rotating star. The value of the departure of temperature is positive (negative) for inclinations lower (higher) than mid-latitudes ($i\sim55^\circ$) and higher its value with higher rotation rate up to the break-up frequency ($\Omega \sim \Omega_{C}$). Then, the departure can be up to $\delta \bar{T}_{\rm eff}(i\sim 0^\circ) \sim 14.5\%$ for pole-on and down to $\delta \bar{T}_{\rm eff}(i\sim 90^\circ) \sim -21.5\%$ for edge-on pure \dScu stars. At mid-latitudes the non-spherical contributions of all structural parameters are the same as a spherically symmetric star \citep{PerezHernandez1999}. \cite{Balona2011} studied the excitation mechanism without taking into account $\Omega$ and $i$. Therefore, we assume that equation~\ref{e:sr1} may be rewritten as
\begin{equation}
\bar{T}_{\rm eff} \propto \nu_{0} \,.
\label{e:sr2}
\end{equation}
Moreover, the mode with highest amplitude can change with time due to any amplitude modulation mechanism \citep[e.g.,][see also Section~\ref{ss:at}]{BarceloForteza2015,Bowman2016}. Taking into account pure \dScu stars only, BF18 use $\nu_{\rm max}$ instead of $\nu_{0}$ as a seismic index,
\begin{equation}
\bar{T}_{\rm eff} \propto \nu_{\rm max} \,, 
\label{e:sr3}
\end{equation}
finding a higher correlation for this scaling relation (see Section~\ref{s:discussion}) and suggesting that gravity-darkening effect may be the cause of the observed dispersion.

\cite{C-D2000} predicted that the age also modifies the excited frequencies due to the increase of the stellar radius. Taking into account hybrid \dScu stars and no gravity-darkening effect, \cite{Bowman2018} suggested that the $T_{\rm eff}$ - $\nu_{0}$ scaling relation should be differentiated for different evolutionary stages.

Here we show how both, gravity-darkening effect and the evolutionary stage, may be the causes of the observed dispersion and how we can use the $\nu_{\rm max}$ as a seismic index. In Section~\ref{s:dSBF}, we explain which data are used and how they are analysed. We present our results for the $\bar{T}_{\rm eff}- \nu_{\rm max}$ scaling relation in Section~\ref{s:results}, including how it changes taking into account different values of surface gravity. In Section~\ref{s:discussion}, we discuss why the evolutionary stage may modify the scaling relation and also the gravity-darkening effect. In Section~\ref{s:corot}, we show the advantages of using the scaling relation to obtain $\bar{T}_{\rm eff}$. Finally, we present our conclusions in Section~\ref{s:conclusion}.

\begin{figure}
\includegraphics[width=\linewidth]{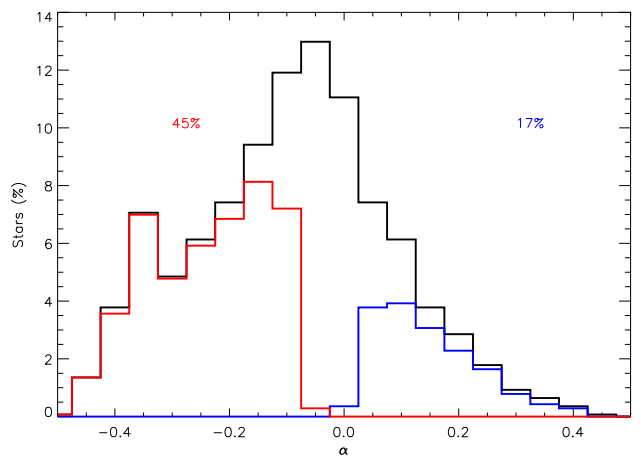}
\caption{Distribution of stars according to their asymmetry (black histogram). Blue (Red) histogram denotes the proportion of stars whose $\nu_{\rm max}$ deviation from the mean frequency of the envelope towards lower (higher) frequencies is significantly higher than the solar case (see text).}
\label{f:alpha}
\end{figure}

\section{Data and analysis}
\label{s:dSBF}

\begin{table*}[!t]
\caption{Parameters of the $\bar{T}_{\rm eff} - \nu_{\rm max}$ relation for each method.}
\label{t:fit}
\centering
\begin{tabular}{c | c c c c c c c}
       & Slope      & Y-intercept & $\sigma$ & r & $P_{u}$ & $N_{in}^{b}$ & $N_{out}^{b}$\\
Method$^{a}$ & (K/$\mu$Hz)& (K)         & (\%)    &   & (\%) & (\%) & (\%) \\
\hline
LFIT$^{1}$ & 2.94 $\pm$ 0.24 & 6980 $\pm$ 50  & 5.82 & 0.424 & $7\times10^{-30}$&  99.3 & 0.7\\
LFIT$^{2}$ & 2.50 $\pm$ 0.10 & 7050 $\pm$ 30  & 5.62 & 0.551 & $6\times10^{-116}$ & 99.4 & 0.6\\
MFIT & 3.34 $\pm$ 0.17 & 6890 $\pm$ 40  & 0.81 & 0.972 & $13\times10^{-19}$& 99.4 & 0.6\\
KFIT & 2.50 $\pm$ 0.55 & 7090 $\pm$ 120 & 1.59 & 0.882 & $4\times10^{-3}$& 99.4 & 0.6\\
\hline \hline
\end{tabular}
\tablefoot{\tablefoottext{a}{See Section~\ref{ss:Tnu}.}\tablefoottext{$b$}{Number of stars in and out of the expected departure of temperature limits taking into account $\Omega \sim \Omega_{C}$ (see text).}\tablefoottext{1}{Values taken from \cite{BarceloForteza2018}.}\tablefoottext{2}{Using the current sample.}}
\end{table*}

Our statistical study needs of a large sample of \dScu stars. For that reason, we analysed a total of 2372 A and F stars with peaks within the typical frequency regime of this type of stars, including those studied in BF18. In addition, we cross-matched our list with that obtained by the Working Group 4 of the \textit{TESS} Asteroseismic Operations Center that excludes well-known or suspected Ap or roAp stars \citep[see][to compare stars from \textit{TESS} first two sectors]{Antoci2019}. This may include Pre-Main Sequence stars, High Amplitude \dScu stars, and slow and fast rotators. We obtained the data from \textit{CoRoT} Sismo-channel: 8 stars \citep{Charpinet2006}; \textit{Kepler} Long (LC) and Short Cadence (SC) light curves: 1124 and 572 stars, respectively \citep{Brown2011}; and \textit{TESS} satellite from sectors 1 to 11 \citep[668 stars;][]{Stassun2019}. We used \textit{Kepler} and \textit{TESS} original data from MAST\footnote{http://stdatu.stsci.edu/kepler/data\_search/search.php \\ http://archive.stsci.edu/tess/bulk\_downloads.html}. Each sector lasts around $\sim$27 days and the number of sectors depend on the position of the star in the sky. Therefore, the maximum duration of the light curve is of $\sim$300 days. The cadence of the studied \textit{TESS} light curves is $\sim$2 minutes and, therefore, the Nyquist frequency is 4167 $\mu$Hz far enough from the typical frequency regime for \dScu stars \citep{Aerts2010}.

\begin{figure}[!t]
\includegraphics[width=\linewidth]{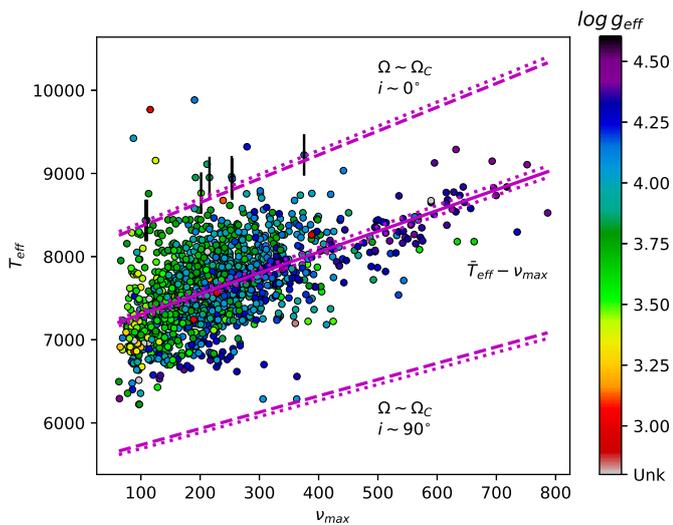}
\caption{Relation between $\nu_{\rm max}$ and $\bar{T}_{\rm eff}$ for \dScu stars (solid line) using LFIT (see text). The color of each star indicates its measured $g_{\rm eff}$ (Unk is for unknown value). Dashed lines mark the limits of the predicted dispersion due to the gravity-darkening effect. All dashed lines represent the estimated error of the linear fit. We show only the error bars for stars in the limit for clarity.}
\label{f:LFIT}
\end{figure}

Using \dScu Basics Finder pipeline \citep[$\delta$SBF hereafter,][and references therein]{BarceloForteza2017}, we characterized their power-spectral structure. Thanks to this method \citep{BarceloForteza2015}, we interpolate the light curve of each star using the information of the subtracted peaks minimizing the effect of gaps and considerably improving the background noise, thereby avoiding spurious effects \citep{Garcia2014}. Finally, this pipeline produces more accurate and precise results in terms of the parameters of the modes. Its reasonably fast computing speed makes this pipeline appropriate for the study of large samples. We also include a superNyquist analysis \citep{Murphy2013} up to 1132 $\mu$Hz only to those light curves with lower Nyquist frequency. This is of importance for those stars observed only with \textit{Kepler} LC in order to properly correct their frequencies and amplitudes. We used the same threshold than in BF18 to study the peaks of the envelope, avoiding hundreds of low amplitude peaks that may be part of the grass \citep[e.g.,][]{Poretti2009,BarceloForteza2017,deFranciscis2019}.

Finally, to test the background improvement of this pipeline for all \textit{TESS} light curves, we used the same method as \cite{Garcia2014} for the \textit{Kepler} data. In addition, we take into account the duty cycle of the observations. The factor of improvement for light curves with duty cycles of 60\% is up to 3. This factor increases up to 14 for duty cycles around 90\%. Moreover, $\nu_{\rm max}$ can be measured with high accuracy (an error up to 5\%) with only 2-d light curves within this range of duty cycles \citep{Moya2018}. Therefore, \textit{TESS} observations are long enough to obtain an accurate value of this seismic index.

\section{Results}
\label{s:results}

\begin{figure*}
\includegraphics[width=0.495\linewidth]{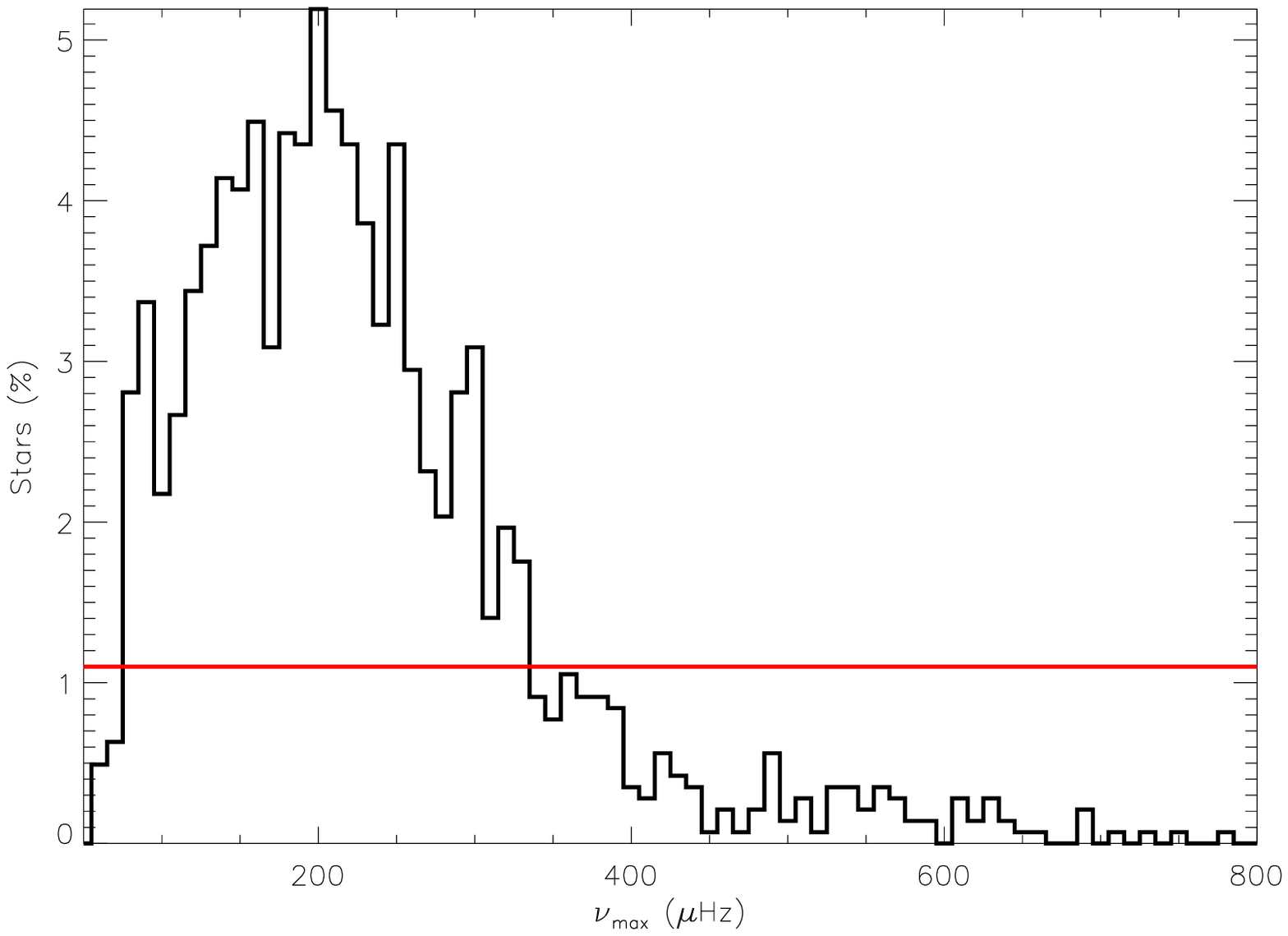}
\includegraphics[width=0.495\linewidth]{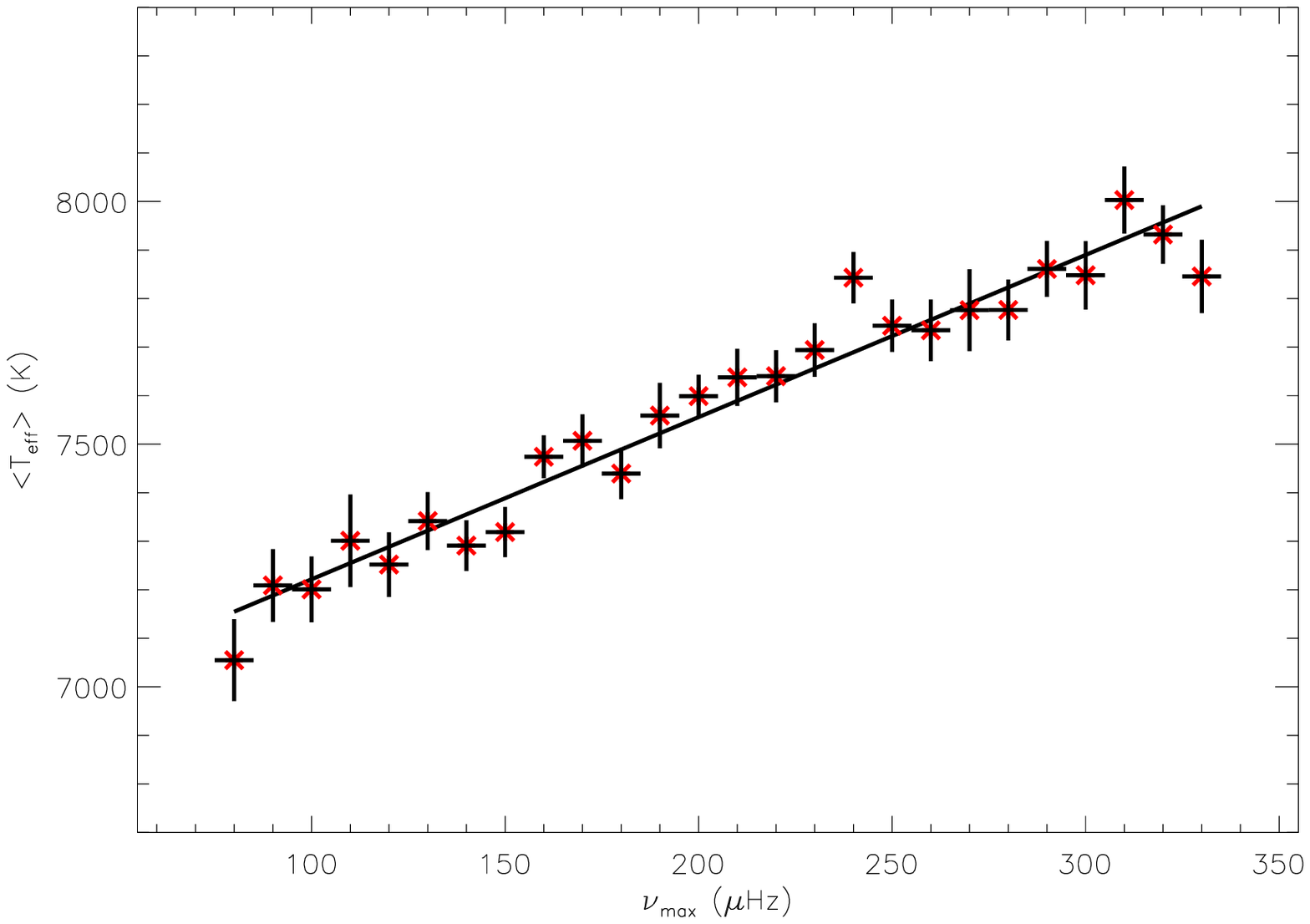}
\caption{\textit{Left panel}: Population of stars for each 10 $\mu$Hz bin of $\nu_{\rm max}$. Red solid line indicates the threshold we used to calculate the MFIT (see text). \textit{Right panel}: Scaling relation found using MFIT.}
\label{f:MFIT}
\end{figure*}

After the analysis, we classified the studied stars in $\delta$~Scuti, $\gamma$~Doradus, and hybrid stars, as explained in \cite{Uytterhoeven2011}. We find that 1442 of the 2372 stars ($\sim$61\%) are \dScu stars, 410 stars ($\sim$17\%) are $\delta$~Sct/$\gamma$~Dor hybrids, 239 stars ($\sim$10\%) are $\gamma$~Dor/$\delta$~Sct hybrids, and 281 stars ($\sim$12\%) are \gDor stars or other kinds of pulsators. We take into account those stars without significant pulsation in the \gDor regime since hybrid stars can have a higher convective efficiency (Uytterhoeven et al. 2011). This is of importance to accomplish all of our assumptions, and only take into account the excitation mechanism of pure \dScu stars oscillations.

Regarding the typical number of peaks in the envelopes, we find between 5 to 37 modes and a mean value of 21 modes (see Fig.~\ref{f:nenv}). Using the acoustic ray dynamics, \cite{Lignieres2009} estimate the number of island modes and chaotic modes of the power spectra of \dScu stars versus the rotation rate. Although its result its only qualitative, we noted that the estimated number of 2-period island modes for a fast-rotating \dScu star is of the same order of magnitude (34 $\pm$ 2 modes).

The observed asymmetry of the envelopes (see Fig.~\ref{f:alpha}) is in agreement with BF18 results. Around 62\% of the envelopes have significantly higher asymmetry than the Sun ($>3\sigma$), $\sim$45\% towards lower frequency modes and $\sim$17\% towards higher frequency modes. The higher number of cases towards lower frequencies may be indicative of the excitation mechanism for this kind of stars.

\subsection{The $\bar{T}_{\rm eff}- \nu_{\rm max}$ scaling relation}
\label{ss:Tnu}

In order to calculate the parameters of this scaling relation we used several techniques. Following the steps in BF18 but including all the stars of the current sample, we made a linear fit between the measured temperatures $T_{\rm eff}$ and $\nu_{\rm max}$ (LFIT, see Table~\ref{t:fit} and Fig.~\ref{f:LFIT}). To test the probability that the relation between these two parameters is not random, we used the Pearson correlation ($r$) and the probability of being uncorrelated \citep[$P_{u}$; i.e.][]{Taylor1997}. This last parameter represents the probability that $N$ measurements of a priori two uncorrelated variables gives a specific Pearson correlation or higher ($|r| \geq  r_{0}$). For example, in the present case, the probability of being uncorrelated with a Pearson correlation coefficient of $R\sim0.55$ is around
\begin{equation}
P_{u} = \frac{2 \Gamma \left( \frac{N-1}{2} \right)}{\sqrt{\pi}\Gamma \left(\frac{N-2}{2} \right)} \int^{1}_{r} \left(1-x^2\right)^{\frac{N-4}{2}} dx \approx 6 \times 10^{-116} \% \, ,
\label{e:Pu}
\end{equation}
where $\Gamma \left( x \right)$ is the gamma function, and $N$ is the number of \dScu stars of the sample. Therefore, we find a statistically significant correlation ($P_{u } \leq 1 $\%). In addition, comparing our results from those of BF18, we noted that the higher number of stars of this kind we add, the lower is the probability than these two parameters are uncorrelated although they have similar dispersion values ($\sigma$; see Table~\ref{t:fit}). As BF18 suggest in their study, this dispersion may be produced by gravity-darkening effect since 99.3\% of the sample lie inside the expected temperature regime.

For the second technique (MFIT), we calculate the mean effective temperature for each 10 $\mu$Hz bin of the $\nu_{\rm max}$. In that way, the different contributions of the departure of temperature ($\delta \bar{T}_{\rm eff}(i)$) produced by the gravity-darkening are cancelled (see Section~\ref{s:intro}). This is only possible if there are a significant amount of stars with a representative amount of different orientations. Then, we only take into account those bins with a population higher than the 1\% of the total amount of stars (see Fig.~\ref{f:MFIT}). Once with those values, we made the fit finding a linear relation with a Pearson coefficient of 0.972. The difference between the parameters obtained from the previous technique can be explained with the shorter range we are forced to take.

The third method (KFIT) requires to know the structural parameters of the \dScu stars. We selected 8 CoRoT \dScu stars from Sismo-channel whose temperature ($T_{\rm eff}$), rotation rate ($\Omega/\Omega_{C}$) and inclination ($i$) has been obtained in other studies (see Table~2 in BF18). Using Eq.~\ref{e:gde}, we can calculate their departure of temperature $\delta \bar{T}_{\rm eff}(i)$ and, finally, its mean effective temperature $\bar{T}_{\rm eff}$. Then, we make the $\bar{T}_{\rm eff} - \nu_{\rm max}$ linear fit. We also find a similar relation than in BF18 but with higher correlation. We noted that the relative differences of mean temperature between all these methods are lower than 5\% (see Section~\ref{s:discussion} for further discussion).

\begin{figure}
\includegraphics[width=\linewidth]{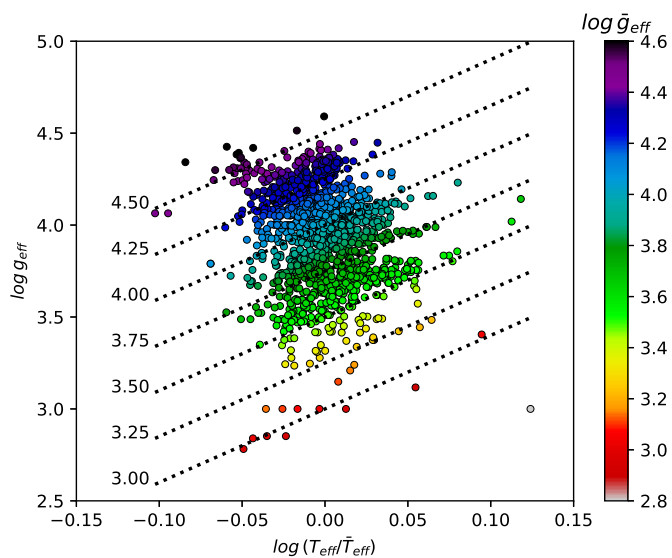}
\caption{Measured surface gravity versus the ratio between measured and mean effective temperature. Different colors for each star indicate their mean surface gravity $\bar{g}_{\rm eff}$. Black dotted lines represent the position of the stars with same $\bar{g}_{\rm eff}$ in the diagram but with different departure of temperature.}
\label{f:Mgeff}
\end{figure}

\subsection{Mean effective gravity}
\label{ss:logg}

\begin{figure*}
\center
\includegraphics[width=0.84\linewidth]{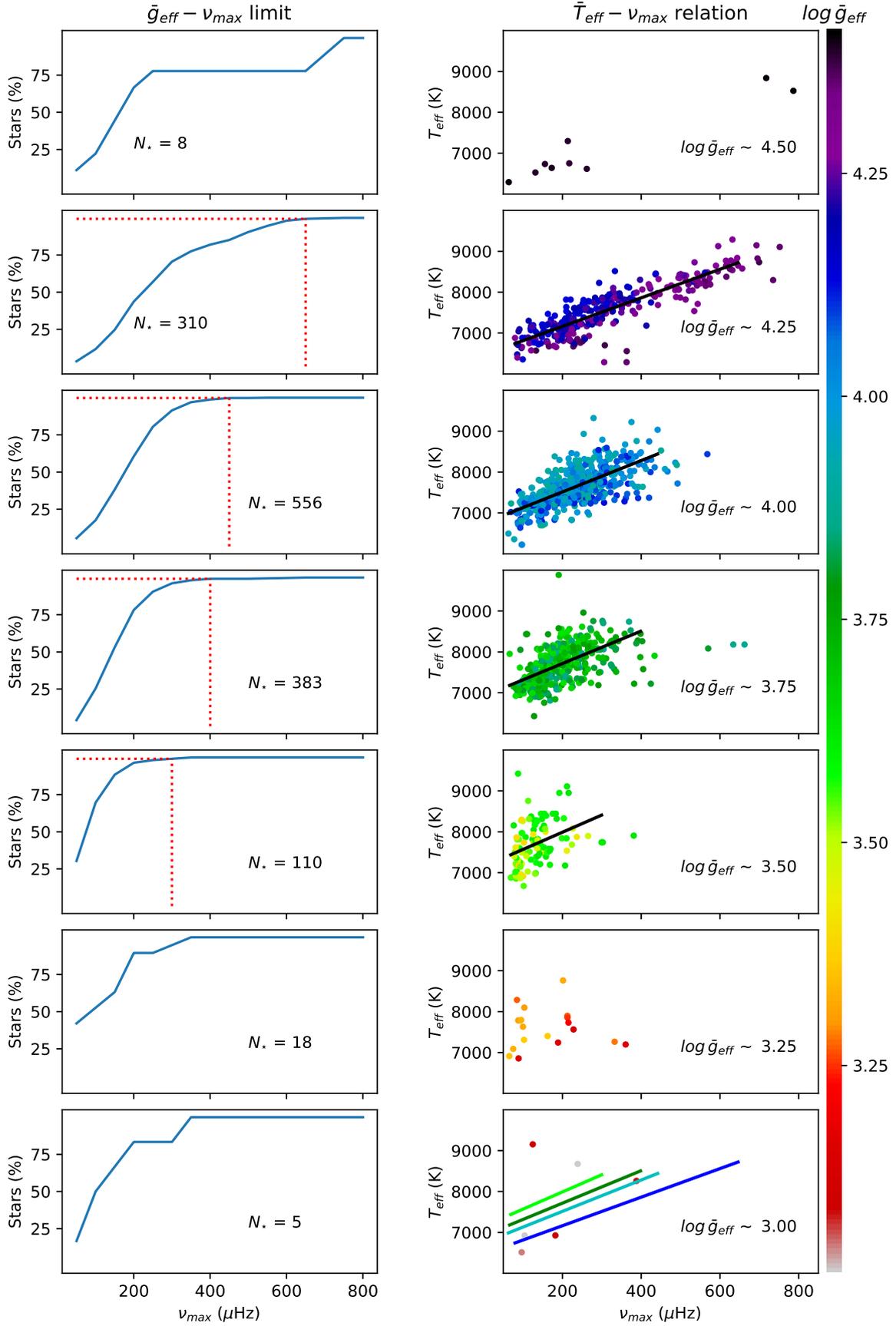}
\caption{\textit{From bottom to top, left panels}: Cumulative histogram of the population of stars per $\nu_{\rm max}$ and higher $\bar{g}_{\rm eff}$. Red dotted lines point to the 99\% of population limit (see text). We indicate the number of stars per group ($N_{\star}$). \textit{From bottom to top, right panels}: Relation between $\nu_{\rm max}$ and $\bar{T}_{\rm eff}$ for \dScu stars of the same group (solid line, see text). The color of each star indicates its  $\bar{g}_{\rm eff}$. \textit{Bottom right panel}: Each colored line represents the scaling relation for each group.}
\label{f:gTu}
\end{figure*}

\begin{table*}
\caption{Parameters of the $\bar{T}_{\rm eff} - \nu_{\rm max}$ relation for each $\bar{g}_{\rm eff}$ group}
\label{t:gfit}
\centering
\begin{tabular}{c | c c c c c c c}
$\log \bar{g}_{\rm eff}$  & Slope      & Y-intercept & $\sigma$ & r & $P_{u}$ & $N_{in}^{\dagger}$ & $N_{out}^{\dagger}$\\
$\pm 0.125$                & (K/$\mu$Hz)& (K)         & (\%)    &    & (\%) & (\%)& (\%)\\
\hline
3.50 & 4.2 $\pm$ 1.1 & 7150 $\pm$ 150  & 6.43 & 0.354 & $15\times10^{-7}$ & 99.1  & 0.9\\
3.75 & 4.0 $\pm$ 0.3 & 6920 $\pm$ 60  & 4.96 & 0.556 & $19\times10^{-35}$ & 99.7 & 0.3\\
4.00 & 3.8 $\pm$ 0.2 & 6750 $\pm$ 40  & 4.25 & 0.680 & $11\times10^{-79}$ & 99.8 & 0.2\\
4.25 & 3.5 $\pm$ 0.1 & 6460 $\pm$ 40 & 3.36 & 0.858 & $4\times10^{-93}$ & 100.0 & 0.0\\
\hline \hline
\end{tabular}
\tablefoot{
\tablefoottext{$\dagger$}{Number of stars in and out of the expected departure of temperature limits taking into account $\Omega \sim \Omega_{C}$ (see text).}}
\end{table*}

\begin{figure}
\includegraphics[width=\linewidth]{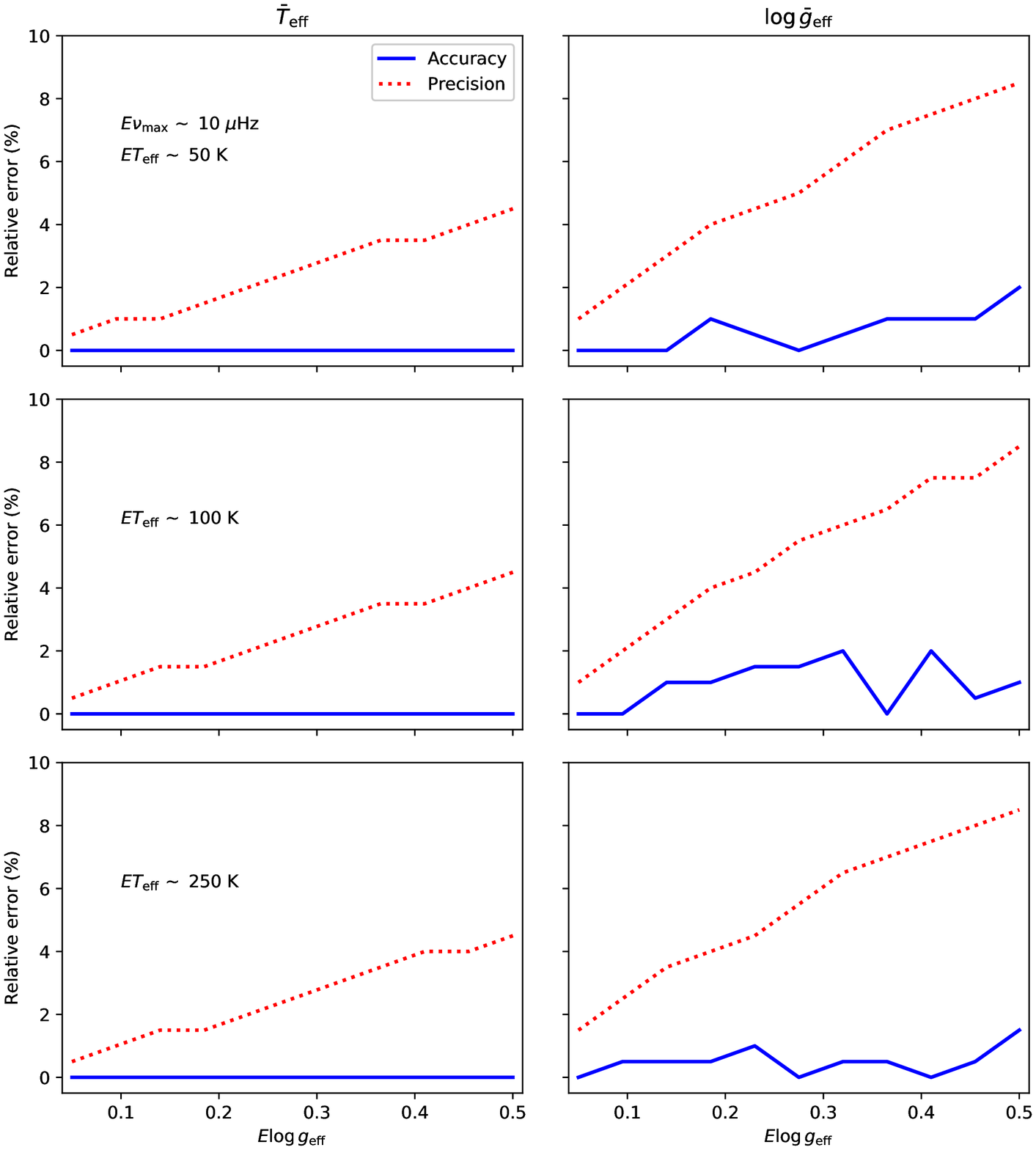}
\caption{Precision and accuracy of our methodology. The relative errors of the mean effective temperature (left) and mean effective surface gravity (right panels) have been calculated for typical photometric temperature errors (from top to bottom) and for different error values of surface gravity.}
\label{f:pE}
\end{figure}

To study the effect of the evolutionary stage in the frequency distribution, \cite{Bowman2018} analysed the power spectra of a large sample of \dScu stars, including hybrids. They separated their sample taking into account the measured effective gravity, $g_{\rm eff}$, considering three different groups: ZAMS ($\log g_{\rm eff} \gtrsim 4.$), MAMS ($3.5 \lesssim \log g_{\rm eff} \lesssim 4.0$), and TAMS ($\log g_{\rm eff} \lesssim 3.5$), for zero-, mid-, and terminal-age main sequence stars, respectively. They conclude that each evolutionary stage should be treated separately.

Here, we do the same excercise but taking into account the gravity-darkening effect. To calculate the mean effective gravity ($\bar{g}_{\rm eff}$), intrinsic to the star, we use von Zeipel's law \citep{vonZeipel1924} 
\begin{equation}
\log \bar{g}_{\rm eff} \approx \log g_{\rm eff}(i) - \frac{4}{\beta} \log \left(\frac{T_{\rm eff}(i)}{\bar{T}_{\rm eff}}\right) \, ,
\label{e:meg}
\end{equation}
where $\beta\sim1$ for stars with fully radiative envelope \citep{Claret1998}; and we obtain $\bar{T}_{\rm eff}$ with LFIT scaling relation. Since we can recover $\bar{g}_{\rm eff}$ for 1390 of our \dScu stars sample (see Fig.~\ref{f:Mgeff}), it is possible to observe if the evolutionary stage affects the $\bar{T}_{\rm eff}-\nu_{\rm max}$ relation. In order to study the dependence of the parameters of the scaling relation with $\bar{g}_{\rm eff}$, we divided our sample in several groups of $\Delta \log \bar{g}_{\rm eff} \sim 0.25$ bins.

First of all, we observe that there is a top limit for $\nu_{\rm max}$ related to the mean surface gravity ($\nu_{d}$). To not to take into account spurious candidates, we define this parameter as the top frequency that contains the 99\% of \dScu stars of its group (see left panels in Fig.~\ref{f:gTu}). We find its dependence with the mean surface gravity with a linear fit,
\begin{equation}
\nu_{d} \sim (224 \pm  26)10^{-4} \bar{g}_{\rm eff} + (240 \pm 30) 
\label{e:maxumax}
\end{equation}
where the frequency is in $\mu$Hz and the mean surface gravity in c.g.s..

Secondly, we made a $\bar{T}_{\rm eff}- \nu_{\rm max}$ linear fit for each mean surface gravity group (see Fig.~\ref{f:gTu}). We have not taken into account those groups with low population and neither those stars with $\nu_{\rm max} > \nu_{d}$. In that way, we only take into account those frequency bins with enough stars to cancel the contribution of the gravity-darkening effect (see Section~\ref{s:intro}). Once with the scaling relation for each $\bar{g}_{\rm eff}$ group (see Table~\ref{t:gfit}), we observed that they change with the mean surface gravity. Then, we calculated the dependence of the slope and the y-intercept with this parameter,
\begin{equation}
\bar{T}_{\rm eff} (\bar{g}_{\rm eff}) \approx \left( a_{1} \bar{g}_{\rm eff} + a_{2} \right) \nu_{\rm max} + \left( a_{3} \bar{g}_{\rm eff} + a_{4} \right) \,.
\label{e:gTnu}
\end{equation}
Once we obtained all parameters ($a_{i}$), we use this improved $\bar{T}_{\rm eff} (\bar{g}_{\rm eff}) - \nu_{\rm max}$ relation in Eq.~\ref{e:meg} to recalculate the mean surface gravity, improving the selection of the respective group for each star. We repeat this process, iterating until the variation of the parameters is negligible ($\delta a_{i}/a_{i} < 10^{-4} \%$). After a few iterations, we obtain the parameters of the improved $\bar{T}_{\rm eff} (\bar{g}_{\rm eff})- \nu_{\rm max}$ scaling relation (see Table~\ref{t:tgnu}) with a probability to be uncorrelated of 8 $\times$ $10^{-212}$\%.

\begin{table}
\caption{Parameters of the improved $\bar{T}_{\rm eff}$($\bar{g}_{\rm eff}$) - $\nu_{\rm max}$ relation}
\label{t:tgnu}
\centering
\begin{tabular}{c | c c }
 & $\nu_{\rm max}^\dagger$  &  $\nu_{0}^\dagger$\\ \hline
$a_{1}$ $\left(\frac{\rm K \, s^{2}}{\rm cm \, \mu Hz}\right)$& -(46 $\pm$ 5) $\times 10^{-6}$  & -(74 $\pm$ 27) $\times 10^{-6}$\\
$a_{2}$ $\left(\frac{\rm K}{\rm \mu Hz}\right)$& 4.30 $\pm$ 0.06 & 3.9 $\pm$ 0.3\\
$a_{3}$ $\left(\frac{\rm K \, s^{2}}{\rm cm}\right)$& (44 $\pm$ 6) $\times 10^{-3}$ & (34 $\pm$ 3) $\times 10^{-3}$\\
$a_{4}$ (K)& 7220 $\pm$ 70  & 7270 $\pm$ 40\\
$\sigma$ (\%) & 1.3 & 1.0\\
 r & 0.701 & 0.669\\
 $P_{u}$ (\%) & $8 \times 10^{-212}$ & $2 \times 10^{-186}$\\
\hline \hline
\end{tabular}
\tablefoot{
\tablefoottext{$\dagger$}{First column shows all parameters of the improved scaling relation (see Eq~\ref{e:gTnu}). The second column presents the same analysis for $\nu_{0}$ instead of $\nu_{\rm max}$.}}
\end{table}

Finally, we find a scaling relation between the frequency at maximum power and two intrinsic parameters of \dScu star structure. To calculate both $\bar{T}_{\rm eff}$ and $\bar{g}_{\rm eff}$ for an individual star, we iterate Eq.~\ref{e:gTnu} and~\ref{e:meg} until these parameters converge to stable values. To test this method, we simulated $\sim10^{9}$ stars with known $\bar{T}_{\rm eff}$, $\bar{g}_{\rm eff}$, $\Omega/\Omega_{C}$, and $i$. We added gaussian noise to their derived $T_{\rm eff}$, $g_{\rm eff}$ and $\nu_{\rm max}$ of the same order of magnitude than expected from \textit{CoRoT}, \textit{Kepler}, and \textit{TESS} catalogues. As we noted in Fig.~\ref{f:pE}, our method allows us to recover the exact value of the mean effective temperature with a precision error up to 4\%. We also find a similar accuracy for the mean effective gravity with a deviation up to 2\% and an error up to 8\%.

\section{Discussion}
\label{s:discussion}

\begin{figure}[!t]
\includegraphics[width=\linewidth]{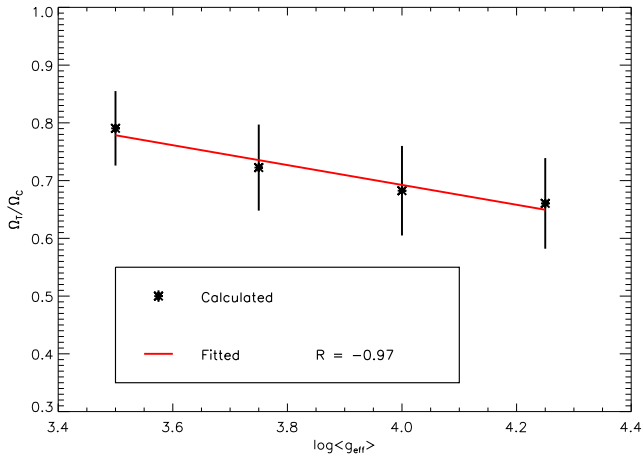}
\caption{Threshold rotation rate per mean surface gravity for \dScu stars. Black asterisks are the calculated values for each group (see text). Red line is a linear fit.}
\label{f:wt-mg}
\end{figure}

We noted that for equal $\bar{T}_{\rm eff}$, the lowest $\bar{g}_{\rm eff}$ \dScu stars excite the lowest frequencies (see bottom  right panel of Fig.~\ref{f:gTu}). Then, older stars should have lower frequency ranges as it is predicted by \cite{C-D2000}. The highest frequency limit, $\sim$800 $\mu$Hz, was already pointed by \cite{Bowman2018} although they only take into account the maximum amplitude peak, $\nu_0$, instead of $\nu_{\rm max}$. To choose a proper parameter to calculate the mean effective temperature is of importance to constrain rotation and inclination for each star (see Section~\ref{ss:iom}). However, BF18 proved that there are not significant differences between the use of both parameters to calculate the scaling relation but $\nu_0$ produce a slightly higher dispersion and lower correlation due to the asymmetry of the envelope (see Section~\ref{ss:at} for further discussion). We repeated the same test with our improved scaling relation (Eq.~\ref{e:gTnu}) finding similar parameters inside $1\sigma$ error (see Table~\ref{t:tgnu}). Therefore, combination frequencies should not affect significantly our results. In addition, there are several studies, both theoretical \citep[e.g.][]{Moskalik1985,Nowakowski2005} and observational \citep[e.g.][]{Breger2014,BarceloForteza2015,Saio2018} suggesting that peaks nearly or equal to combination of other frequencies may be resonantly-excited modes. In that way, these modes should be taken into account to calculate $\nu_{\rm max}$.

Another phenomenon we observe is a higher dispersion of temperatures for lower $\bar{g}_{\rm eff}$ groups ($\sigma$, see Table~\ref{t:fit} and Fig.~\ref{f:gTu}). The gravity-darkening effect depends on the ratio between centrifugal and gravity forces, $\epsilon^2$ (see Eq.~\ref{e:gde}). Rewritting Equation~\ref{e:eps2} as
\begin{equation}
\epsilon^2 = \frac{\Omega^2 R}{\bar{g}_{\rm eff}} \propto \frac{\Omega^2}{\bar{\rho}}\, ,
\label{e:eps2g}
\end{equation}
we noted that a higher rotation is required for more dense stars to have the same $\epsilon^2$, i.e., the same departure of temperature. Combining the gravity-darkening effect and the stellar evolution theory, we may explain the behaviour of temperature dispersion since radius increase with age \citep{C-D2000}. Assuming that the observed dispersion ($\sigma$) is produced by the gravity-darkening effect, we can define the threshold rotation rate ($\Omega_{T}$) as the minimum rotation needed to observe a departure of temperature equal to $\sigma$. We can calculate the minimum rotation rate of a particular departure of temperature assuming a pole-on or equator-on star since intermediate values of inclination require higher values of rotation (see Section~1 and BF18). We use a numerical technique to calculate this parameter (see Section~\ref{ss:iom} for further details). Our results suggest that $\Omega_{T}/\Omega_{C}$ decrease with the mean surface gravity (see Fig.~\ref{f:wt-mg}) with the form
\begin{equation}
\frac{\Omega_{T}}{\Omega_{C}} \approx -(0.17 \pm 0.03) \log \bar{g}_{\rm eff} + (1.38 \pm 0.11) \,, 
\label{e:wg}
\end{equation}
and, therefore, increase with age. This effect does not mean that rotation should increase with age \citep[contrary to gyrochronology predictions; e.g.,][]{Soderblom2010}, but it might decrease less than density. In any case, all the values of $\Omega_{T}/\Omega_{C}$ are in agreement with the great fraction of fast rotators for A-type stars found by \cite{Royer2007}: $\Omega \gtrsim 0.5 \Omega_{C}$.

\begin{figure}
\includegraphics[width=\linewidth]{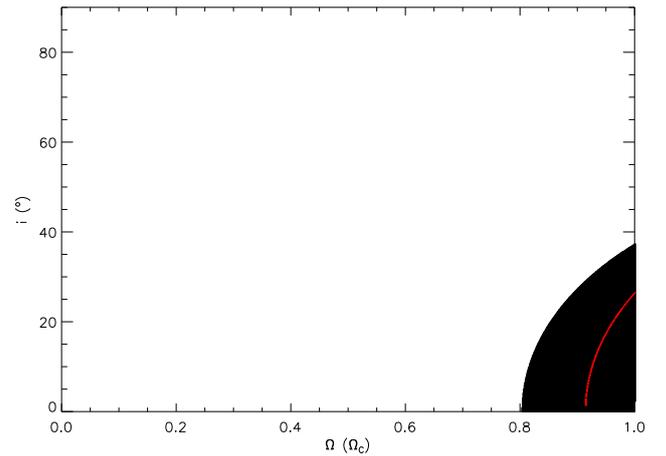}
\caption{$i-\Omega$ map of the \textit{Kepler} \dScu star KIC~11823661. Red points are the models with the exact value of observed departure of temperature. Black points represent those models that take into account the error bars.}
\label{f:mapir}
\end{figure}

\subsection{$i-\Omega$ maps}
\label{ss:iom}

\begin{figure*}
\center
\includegraphics[width=0.9\linewidth]{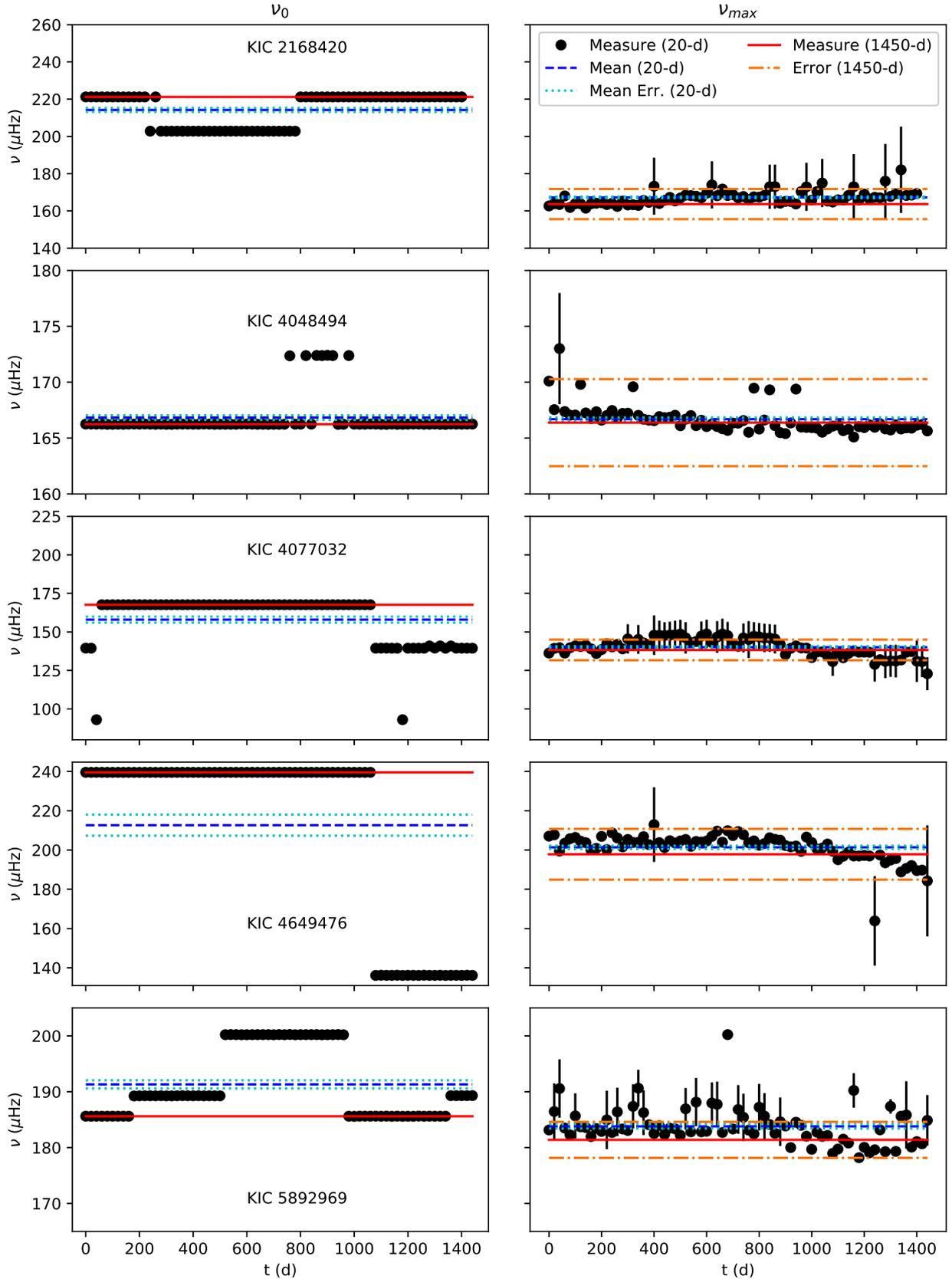}
\caption{Frequency of the highest amplitude peak ($\nu_{0}$, left panels) and frequency at maximum power ($\nu_{\rm max}$, right panels) with time for five pure \dScu stars with detected RMC (one per row; see text). Black circles represent the measurements of each parameters for 20-day segments of the entire light curve. Blue dashed line is the mean value of all 20-d measurements and blue dashed-dotted lines are their error. Red line is the measurement of each parameter for a 1450-day light curve and red dotted lines are their error. The error bars for 20-d measurements of $\nu_{0}$ are smaller than the symbol. For clarity reasons, we only plotted the error bars for 20-d measurements of $\nu_{\rm max}$ for these outside of the 1450-d measure error.}
\label{f:cumax}
\end{figure*}

\begin{figure*}[!t]
\center
\includegraphics[width=0.9\linewidth]{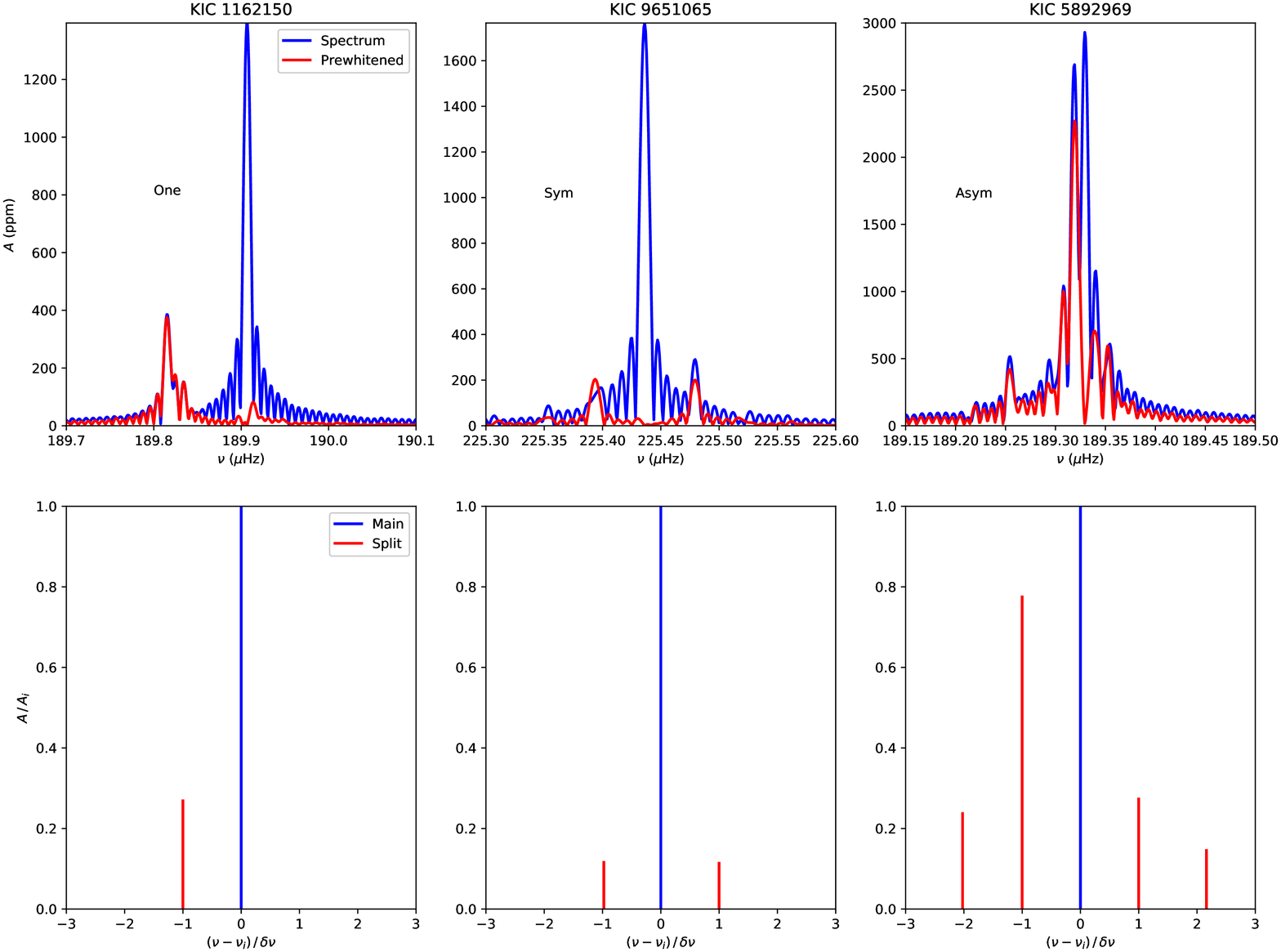}
\caption{Multiplets with one (left), symmetric (middle), and asymmetric (right panels) sidelobes. Top panels: Detail of the original 10-point oversampled power spectrum (blue) and also after extracting the central peak (red). Bottom panels: Detected peaks using $\delta$SBF taking into account the frequency shift between split peaks and their amplitude ratios. Blue points to the central peak and red points to split peaks.}
\label{f:multi}
\end{figure*}

BF18 calculated the minimum rotation rate ($\Omega_{\rm min}$) for $\sim$700 pure \dScu stars. This parameter can be obtained using Eq.~\ref{e:gde} and assuming the star is pole-on or equator-on. Furthermore, the limits of the inclination from the line of sight ($i$) can also been obtained assuming $\Omega \sim \Omega_C$. But, the observed departure of temperature should be higher than the relative error of the measurement
\begin{equation}
\delta \bar{T}_{\rm eff,obs} > ET_{\rm eff} / \bar{T}_{\rm eff} \, , 
\label{e:dtobs}
\end{equation}
to avoid the $i-\Omega$ degeneracy zone (see Fig 5 and 6 in BF18). This zone does not allow us to use this technique to differentiate between a moderate or slow rotator with any inclination from a extreme rotator with an inclination close to the mid-latitude ($i\sim 55^\circ$). In that case a deeper study is needed \citep[e.g.,][]{Poretti2009,GarciaHernandez2013,Escorza2016,BarceloForteza2017}.

We used a different technique to obtain a map with all possible combinations of $i-\Omega$. This method consits to simulate around one million of stars with different $i-\Omega$ values and only select those which fulfil the observed departure of temperarure $\delta \bar{T}_{\rm eff,obs}$ (see Fig.~\ref{f:mapir}).

To calculate the correct $\delta \bar{T}_{\rm eff,obs}$, we take into account the improved scaling relation (Eq.~\ref{e:gTnu}). But it is not always possible since we may not know the measured $g_{\rm eff}$. In that case we use the LFIT scaling relation. The limits of rotation and inclination for the stars of our sample (see Table~A.1), including $i-\Omega$ maps, are only available in electronic form.

In this way, we find only five \dScu stars are out of the expected regime of temperature for fast rotators: $\delta T_{\rm eff,obs} \lesssim -21.5 \% $ or $\delta T_{\rm eff,obs} \gtrsim 14.5 \% $. The other 4 outsiders have an unknown value of surface gravity. These stars may be (pre-)Extremely Low Mass stars since they have similar frequency ranges and similar or higher surface gravity \citep{SanchezArias2018}.

\subsection{Mode variations with time}
\label{ss:at}

There are several reasons to use $\nu_{\rm max}$ instead of $\nu_0$ apart from its lower correlation. First of all, the visibility of the highest amplitude mode depends of the point of view of the observer \citep[see][]{Lignieres2009}. Secondly, the highest amplitude mode is not fixed, i.e., the amplitudes can change with time and other modes can become the highest amplitude mode \citep[i.e.,][]{Handler1998,Breger2000a,BarceloForteza2015}. This is not the case for $\nu_{\rm max}$ that remains approximately constant during the cyclic changes (see Fig~\ref{f:cumax}). To study the variability of $\nu_{0}$ and $\nu_{\rm max}$ with time for stars with detected cyclic variations, we used $\delta$SBF pipeline for each 20-d segments and we compared the results with these of the entire light curve. To find a sample of stars with cyclic variations, we studied the power spectrum of the entire light curve for all pure \dScu stars of our sample. There, the variations in the parameters of a mode are observed as split peaks of this mode \citep[e.g.,][see also Fig.~\ref{f:multi}]{Moskalik1985,ShK2012}, i.e., a multiplet. The frequency shift between peaks, the ratio of amplitudes, and the symmetry of the multiplet may indicate the nature of the variation. On one hand, a symmetric multiplet could indicate a superNyquist frequency mode \citep{Murphy2013} or a binarity nature of the system \citep[e.g.,][]{ShK2012,Murphy2014}. On the other hand an asymmetric multiplet can indicate a cyclic variation such as resonant mode coupling \citep[RMC;][]{Moskalik1985,BarceloForteza2015} or, in extreme cases, may suggest a definitive change in the stellar structure \citep{Bowman2014}. Fig.~\ref{f:cumax} show the variation of $\nu_{0}$ and $\nu_{\rm max}$ for 5 pure \dScu stars with detected variations in some of their peaks in agreement with RMC. The sharp changes of $\nu_{0}$ with time can be compared with $\nu_{\rm max}$ 20-d measurements. In fact, the $\nu_{0}$ mean of the 20-d light curves is not in agreement with the value obtained with the entire light curve in all tested cases. This is not the case for $\nu_{\rm max}$ since both values are equal within errors. Resonances seem not to modify $\nu_{\rm max}$ at least in the long term.

\begin{figure}
\includegraphics[width=\linewidth]{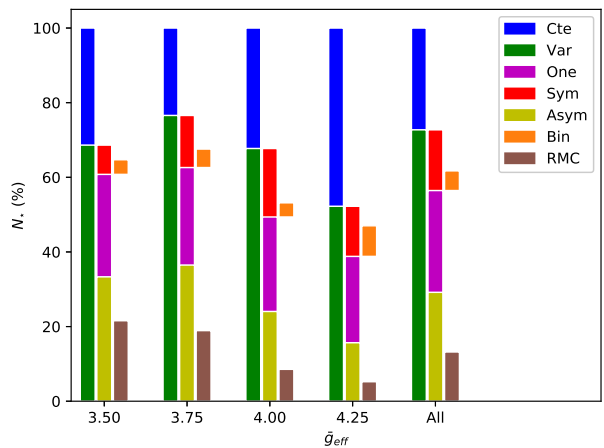}
\caption{Proportion of pure \dScu stars with constant (Cte) or variable (Var) modes for all the sample and for different evolutionary stage. This last kind can be differentiated with the shape of the multiplet: one sidelobe (One), symmetric (Sym) or asymmetric (Asym) sidelobes. (Bin) indicates the proportion of binary stars detected with this method (see text). RMC indicates the proportion of this kind of stars that show multiplets affected by resonant mode coupling (see text).}
\label{f:histovar}
\end{figure}

In this study, we also find that only 27\% of pure \dScu stars have constant amplitude peaks (see Fig.\ref{f:histovar}). We recover the same proportion of stars with constant modes obtained by \cite{Bowman2016} if we take into account hybrid stars (38\%). We observed that the other 73\% of the pure \dScu stars have modes with detected amplitude and/or phase variations. Looking to these phenomena for each $\bar{g}_{\rm eff}$ group, we observe significant differences. The lower the surface gravity, the larger fraction of \dScu stars with detected variable modes (from 52\% to 76\%), especially for those with asymmetric multiplets (from 16\% to 35\%), including these candidates to have RMC (from 5\% to 20\%). Moreover, these candidates with extrinsic causes of variation (superNyquist frequencies or binarity) are approximately constant with surface gravity ($\sim$16\%). Finally, the proportion of stars observed with multiplets of only one detected sidelobe is approximately constant too ($\sim$27\%). We also added this analysis star by star in Table~A.1.

Our results suggest that the evolutionary stage seems to favour resonances towards older ages. This is in agreement with the increase of the g-mode frequencies with age \citep{C-D2000} and its interaction with p-modes. In addition, the transition stages may be observed as permanent changes in the power spectra of the stars due to their restructuring.

\section{Advantages of $\bar{T}_{\rm eff}$}
\label{s:corot}

\begin{figure}
\includegraphics[width=\linewidth]{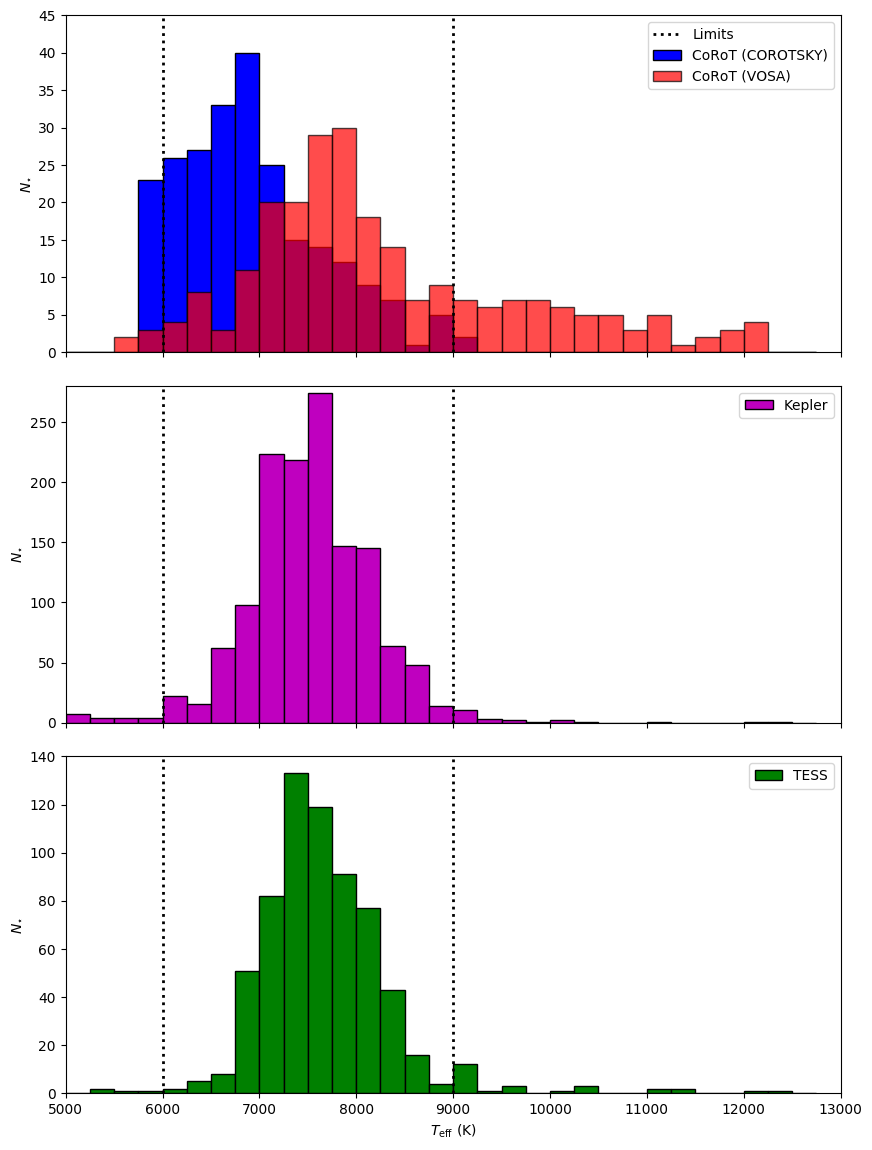}
\caption{Top panel: Histogram of temperatures for 239 \dScu star candidates observed by \textit{CoRoT}. Blue bars point to values from COROTSKY database \citep{Charpinet2006} and red bars point to these obtained with VOSA (see text). Dotted lines represent the temperature limits for \dScu stars \citep{Uytterhoeven2011}. Middle and bottom panels: Same as top panel for stars of our main sample observed by \textit{Kepler} and \textit{TESS}, respectively.}
\label{f:hT}
\end{figure}

Once with the parameters of the scaling relation, we can use Eq.~\ref{e:gTnu} to obtain the mean effective temperature of other pure \dScu stars. We analysed the power spectra of 239 \dScu candidates observed by CoRoT Exo-channel \citep{Debosscher2009}, calculating their $\bar{T}_{\rm eff}$ for 174 of them (see Table~A.2 only in the electronic form). The $T_{\rm eff}$ of these stars were estimated by fitting their spectral energy distribution (SED) to a grid of theoretical models \citep[Kurucz,][]{Castelli1997} using the Virtual Observatory tool \citep[VOSA,][]{Bayo2008}. Extinction was left as a free parameter in the SED fitting process, ranging from zero to the value obtained from the NASA/IPAC Galactic Dust Reddening and Extinction service\footnote{https://irsa.ipac.caltech.edu/applications/DUST/} using \citep{Schlafly2011}

Figure~\ref{f:hT} compares the effective temperatures obtained with VOSA (red bars) with those available in the COROTSKY Database \citep[blue bars;][]{Charpinet2006}. We can see how the assumption of no extinction for the majority of
the objects in COROTSKY leads to an underestimation of the temperatures. This effect may cause a misclassification since cool \dScu stars are discarded and other kind of hot pulsators are included.

Moreover, different models based on different physical properties, may produce results with discrepancies up to the same order of magnitude \citep{Sarro2013}. For example, rotation can modify the observed colors \citep{Collins1985} with the consequent impact on temperature. In contrast, $\bar{T}_{\rm eff}$ seem not to depend on these parameters. Therefore, we can conclude that the scaling relation based on $\nu_{\rm max}$ allow us to characterize \dScu stars independently of rotation and also extrinsic parameters of the star such as extinction.

\section{Conclusions}
\label{s:conclusion}

The relation between the power-spectral structure and the structural parameters for \dScu stars has been a long-standing debate, especially since the beginning of large surveys thanks to space telescopes \citep[e.g.,][]{Balona2011,Moya2017}. In this work, we have studied the oscillation spectra of 2372 A-F pulsating stars observed by \textit{CoRoT}, \textit{Kepler} \& \textit{TESS}. From them, 1442 were pure \dScu stars. Once characterized their power spectra, we obtained the empirical scaling relation between the frequency at maximum power, the mean effective temperature, and the mean surface gravity (Eq~\ref{e:gTnu}). This is in agreement with the predicted frequency distribution for $\kappa$-mechanism \citep{Dziembowski1977} since we detected higher frequency modes for higher temperature \dScu stars. In fact, our relation is similar to that found by \cite{BarceloForteza2018}. We also observed that old stars with low surface gravity present lower frequency ranges (see Eq.~\ref{e:maxumax} and Fig.~\ref{f:gTu}), just as opposite than young \dScu stars. This is in agreement with predictions too \citep{C-D2000}. Therefore, the evolutionary stage affects $\bar{T}_{\rm eff}-\nu_{\rm max}$ relation and it must be taken into account to find the intrinsic parameters of these kind of stars ($\bar{T}_{\rm eff},\bar{g}_{\rm eff}$).

Photometric and spectroscopic techniques to measure the temperature ($T_{\rm eff}$) and the surface gravity ($g_{\rm eff}$) may be significantly affected by gravity-darkening effect \citep{vonZeipel1924}. Then, we have developed a methodology to correct such effect by iterating Eqs.~\ref{e:meg} and~\ref{e:gTnu} until convergence. Gravity-darkening may also explain the observed dispersion of the scaling relation ($\sigma$) and its decrease with $\log{\bar{g}_{\rm eff}}$ (see Table~\ref{t:gfit}). Our results suggest that ageing stars may decrease its rotation slowly than its density, making them closer to its break-up frequency. Thanks to the departure of temperature of each individual star (Eq.~\ref{e:gde}), we have delimited rotation and inclination from the line of sight, especially for fast rotators. In that way, it would be possible to correct their position in the HR~Diagram and then improve age determination using isochrone fitting \citep[e.g.][]{Michel1999,FoxMachado2006}. In addition, since \dScu stars are used as standard candles \citep[e.g.][]{McNamara2011,Ziaali2019}, it would be feasible to improve the distance determination to globular clusters and other galaxies. Moreover, exoplanetary research may benefit from our method since the calculation of the habitable zone depends on stellar parameters such as $\bar{T}_{\rm eff}$ \citep{Kopparapu2014}.

In conclusion, we suggest the frequency at maximum power ($\nu_{\rm max}$; see Eq.~\ref{e:numax}) as a seismic index since it is a proper indicative of the mean temperature and surface gravity of the star. It is independent of rotation and extrinsic parameters such as inclination or extinction. Furthermore, $\nu_{\rm max}$ is not as affected as the highest amplitude mode ($\nu_{0}$) by resonances. This property is especially useful for older stars since age benefits the interaction between modes (see Fig.~\ref{f:histovar}). Finally, $\nu_{\rm max}$ variation may indicate the restructuring of the stars and their power spectra between different transition stages.

\paragraph{}
\begin{acknowledgements}
Comments from J.A.~Caballero are gratefully acknowledged. Authors wish to thank the referee for useful suggestions that improved the paper. We also thank the \textit{CoRoT}, \textit{Kepler}, and \textit{TESS} Teams whose efforts made these results possible. The \textit{CoRoT} space mission has been developed and was operated by \textit{CNES}, with contributions from Austria, Belgium, Brazil, ESA (RSSD and Science Program), Germany and Spain. Funding for \textit{Kepler}'s Discovery mission is provided by NASA’s Science Mission Directorate. Funding for the TESS mission is provided by the NASA Explorer Program. Authors acknowledge the effort made by TASOC WG4 that helped us in our target selection. This publication makes use of VOSA, developed under the Spanish Virtual Observatory project supported by the Spanish MICIU through grant AyA2017-84089. VOSA has been partially updated by using funding from the European Union's Horizon 2020 Research and Innovation Programme, under Grant Agreement No 776403 (EXOPLANETS-A). SBF and DB has received financial support from the Spanish State Research Agency (AEI) Projects No.ESP2017-87676-C5-1-R and No. MDM-2017-0737 Unidad de Excelencia “María de Maeztu”- Centro de Astrobiología (INTA-CSIC). AM acknowledges funding from the European Union’s Horizon 2020 research and innovation program under the Marie Sklodowska-Curie grant agreement No 749962 (project THOT). SMR acknowledges financial support from the State Agency for Research of the Spanish MICIU through the "Center of Excellence Severo Ochoa" award to the Instituto de Astrof\'isica de Andaluc\'ia (SEV-2017-0709). JCS and AGH acknowledges funding support from Spanish public funds (including FEDER fonds) for research under project ESP2017-87676-C5-2-R and ESP2017-87676-C5-5-R. JCS also acknowledges support from project RYC-2012-09913 under the "Ram\'on y Cajal" program of the Spanish Ministry of Science and Education. AGH acknowledges support from "Universidad de Granada" under project E-FQM-041-UGR18 from "Programa Operativo FEDER 2014-2020" programme by "Junta de Andaluc\'{\i}a" regional Government.

\end{acknowledgements}

\bibliographystyle{aa}
\bibliography{tcb}

\end{document}